\documentclass[final,3p,times]{elsarticle}

\usepackage{amssymb}
\usepackage{amsmath}
\usepackage{natbib}

\usepackage{amscd,amsthm,verbatim,color,fancyhdr,mathrsfs}
\usepackage{graphicx}
\usepackage{subcaption}
\usepackage{turnstile}
\usepackage{qcircuit}
\usepackage{float}
\usepackage[colorlinks,linkcolor=blue,anchorcolor=red,citecolor=red]{hyperref}
\usepackage{cleveref}
\usepackage{tikz}
\usepackage{lmodern}
\usepackage{booktabs}

\newtheorem{thm}{Theorem}
\newtheorem{lem}[thm]{Lemma}
\newtheorem{cor}[thm]{Corollary}
\newtheorem{prop}[thm]{Proposition}
\newdefinition{rmk}{Remark}
\newdefinition{ex}{Example}
\newtheorem{defn}[thm]{Definition}
\newproof{pf}{Proof}

\newcommand{\A}{\boldsymbol A}

\newcommand{\T}{\boldsymbol T}

\newcommand{\U}{\boldsymbol U}
\newcommand{\D}{\boldsymbol{D}}

\newcommand{\C}{\boldsymbol C}

\newcommand{\y}{\boldsymbol y}
\newcommand{\B}{\boldsymbol B}

\newcommand{\I}{\boldsymbol I}
\newcommand{\0}{\boldsymbol O}

\newcommand{\vv}{\boldsymbol v}
\newcommand{\V}{\boldsymbol V}

\newcommand{\F}{\boldsymbol F}
\newcommand{\HH}{\boldsymbol H}

\newcommand{\X}{\boldsymbol X}

\newcommand{\PP}{\boldsymbol P}
\newcommand{\aaa}{\boldsymbol a}
\newcommand{\cc}{\boldsymbol c}

\begin{document}

\begin{frontmatter}



\title{Products between block-encodings}


\author[fd]{Dekuan Dong} 
\author[fd,shkey]{Yingzhou Li} 
\author[fd]{Jungong Xue}
\affiliation[fd]{organization={School of Mathematical Science, Fudan University},
            country={China}}

\affiliation[shkey]{organization={Shangehai Key Laboratory for Contemporary Applied Mathematics},
            country={China}}

\begin{abstract}
Block-encoding is a standard framework for embedding matrices into unitary operators in quantum algorithms. Efficient implementation of products between block-encoded matrices is crucial for applications such as Hamiltonian simulation and quantum linear algebra. We present resource-efficient methods for matrix-matrix, Kronecker, and Hadamard products between block-encodings that apply to rectangular matrices of arbitrary dimensions. Our constructions significantly reduce the number of ancilla qubits, achieving exponential qubit savings for sequences of matrix-matrix multiplications, with a moderate increase in gate complexity. These product operations also enable more complex block-encodings, including a compression gadget for time-dependent Hamiltonian simulation and matrices represented as sums of Kronecker products, each with improved resource requirements.
\end{abstract}




\end{frontmatter}



\section{Introduction}
\label{sec1}
Quantum computing holds the potential to solve certain computational problems far more efficiently than classical algorithms. This promise is especially evident in areas dominated by linear algebraic operations, such as Hamiltonian simulation \cite{10.1109/FOCS.2015.54,low2019hamiltoniansimulationinteractionpicture}, solving systems of linear equations \cite{PhysRevLett.103.150502,doi:10.1137/130912839,PRXQuantum.3.040303}, quantum machine learning \cite{biamonte2017quantum}, and quantum optimization \cite{farhi2014quantumapproximateoptimizationalgorithm, RevModPhys.90.015002}. A central reason for this advantage is that quantum computers can naturally manipulate high-dimensional vector spaces, provided that the underlying linear operator can be accessed efficiently. 

The block-encoding framework, formalized in \cite{10.1145/3313276.3316366}, has emerged as a powerful method for embedding matrices into unitary operations, thereby enabling quantum algorithms to operate on non-unitary matrices. The block-encoding technique embeds a matrix $\A\in \mathbb C^{2^{\scriptstyle{n}}\times 2^{\scriptstyle{n}}}$ into the top-left block of a larger unitary operator $\U_{\A}$, allowing quantum circuits to access and manipulate $\A$ indirectly while preserving unitary property. Formally, a unitary $\U_{\A}\in \mathbb C^{2^{\scriptstyle{n + a}}\times 2^{\scriptstyle{n+a}}}$ is called an $(\alpha, a, \varepsilon)$-block-encoding of $\A$ if 
\[\left\| \A - \alpha \left(\langle 0^a|\otimes \I\right) \U_{\A}\left(|0^a\rangle \otimes \I\right)\right\| \le \varepsilon,\]
where $a$ ancilla qubits are initialized and postselected in the $|0^a\rangle$ state. By representing a matrix as a block of a larger unitary, block-encoding provides a uniform language for quantum linear algebra, streamlining both the analysis and design of algorithms. Consequently, block-encoding has become a cornerstone technique in the development of efficient quantum algorithms. The technique of block-encoding enables efficient implementation of matrix functions \cite{10.1145/3313276.3316366}, quantum linear system solvers \cite{PRXQuantum.3.040303}, and Hamiltonian simulation techniques \cite{low2019hamiltoniansimulationinteractionpicture, PhysRevLett.118.010501}. Consequently, a growing body of work has focused on constructing block-encodings for individual matrices or matrix classes \cite{camps2024explicit,Sunderhauf2024blockencoding,LIU2025102480,9951292,guseynov2025efficientexplicitgateconstruction,Nguyen2022blockencodingdense,yang2025dictionarybasedblockencodingsparse}. 

In many applications, however, the matrices of interest do not appear in isolation but rather as compositions of simpler matrices. For instance, Dyson-series based Hamiltonian simulation requires ordered products of Hamiltonian terms \cite{low2019hamiltoniansimulationinteractionpicture}; quantum differential equation algorithms often involve structured matrices with Kronecker product forms \cite{JIN2025114138,9618807,Dong2025quantumalgorithm}. Thus, the task of constructing block-encodings for compositions of matrices arises frequently and is indispensable for a broad range of quantum algorithms. Several foundational matrix operations have been shown to be efficiently implementable between block-encodings:
\begin{itemize}
    \item \textbf{Matrix-matrix multiplication:} Given an $(\alpha, a, \varepsilon)$-block-encoding of $\A$ and a $(\beta, b, \delta)$-block-encoding of $\B$, one can construct an $(\alpha\beta, a + b, \alpha \varepsilon + \beta\delta)$-block-encoding of $\A\B$ as described in \cite{10.1145/3313276.3316366}. 
    \item \textbf{Kronecker product:} If $\A$ and $\B$ are block-encoded, their Kronecker product $\A\otimes \B$ can be block-encoded by taking the tensor product of the two block-encodings and projecting onto the combined ancilla state \cite{PhysRevA.102.052411}.
    \item \textbf{Matrix Hadamard product:} Since the entries required for the Hadamard product are contained within the Kronecker product, it can be derived by first constructing the Kronecker product and then applying a proper permutation to reposition the relevant elements \cite{guo2024quantumlinearalgebraneed}.  
    \item \textbf{Linear combinations:} Using the Linear Combination of Unitaries (LCU) technique, weighted sums of block-encoded matrices can be implemented with logarithmic overhead in the number of terms \cite{10.1145/3313276.3316366}.
    \item \textbf{Inversion and matrix functions:} The technique called quantum singular value transformation (QSVT) enables efficient implementation of functions of block-encoded normal matrices, including $\A^{-1}$ and $\exp(\imath \A t)$ \cite{10.1145/3313276.3316366,PhysRevLett.118.010501}. 
\end{itemize}

This paper is devoted to the problem of implementing products between block-encodings, including the Kronecker product, the Hadamard product, and the matrix-matrix product, in a qubit- and gate-efficient manner. Our goal is to exploit matrix structures to design product constructions that reduce circuit complexity and ancillary qubit requirements, thereby extending the practicality of block-encoding based algorithms on quantum devices. Our main contributions are as follows:
\begin{itemize}
    \item We address the more general setting where the two matrices to be multiplied are not necessarily square, and their dimensions are not restricted to powers of two, thereby broadening the applicability of block-encoded matrix operations.
    \item We propose a new method for implementing matrix-matrix products that significantly reduces the number of required ancilla qubits, at the cost of a modest increase in gate complexity. The qubit savings become particularly significant when computing the product of a sequence of matrices, where the required number of ancilla qubits can be reduced exponentially.   
    \item Compared to the Kronecker product implementation in \cite{PhysRevA.102.052411}, we replace the SWAP gates with two CNOT gates, achieving a reduction in gate complexity. Extending the same qubit-efficient principle used in our matrix-matrix product construction, we further provide a qubit-efficient method for implementing the Kronecker product.
    \item We rederive the Hadamard product implementation between block-encodings from \citep{guo2024quantumlinearalgebraneed} and extend it to construct convolution operations and vectorization operators, thereby enriching the toolbox of linear operations that can be efficiently realized within block-encodings framework. 
    \item We apply the idea of qubit-efficient implementation of matrix-matrix products to construct a compression gadget for time-dependent Hamiltonian simulation \cite{low2019hamiltoniansimulationinteractionpicture}, which has broad applications including non-unitary dynamic simulation \cite{JIN2024112707, an2023quantumalgorithmlinearnonunitary,Berry2024quantumalgorithm} and quantum linear system algorithms \cite{10.1145/3498331}. Compared to the existing method, our construction is more concise and eliminates redundant control operations. 
    \item By combining the Kronecker product of block-encodings with the LCU framework, matrices represented as sums of Kronecker products can be efficiently block-encoded. The benefit of utilizing such a structure is exemplified by the adjacent matrix of an extended binary tree \cite{camps2024explicit}, where our approach naturally yields a more efficient implementation.
\end{itemize}
The remainder of this paper is organized as follows. In \Cref{sec:2}, we present our constructions for block-encodings of the Kronecker product, the matrix-matrix multiplication, and the Hadamard product. Implementations of convolution operators and vectorization operators are also discussed. In \Cref{sec:3}, we give several applications of matrix products between block-encodings, including the relevance to time-dependent Hamiltonian simulations and the block-encoding of matrices which can be written as a sum of Kronecker products. Finally, \Cref{sec:4} provides concluding remarks and directions for future research.
\section{Implementation of products}\label{sec:2}
In this section, we describe the implementation of several types of products between block-encodings. We begin with the Kronecker product, which naturally aligns with the tensor-product structure of quantum computing and can therefore be realized efficiently. Then, we introduce a qubit-efficient method for implementing matrix-matrix products, which substantially reduces the required number of ancilla qubits; the same idea can also be applied to the Kronecker product. Finally, we present an explicit construction of the permutation matrix that extracts the entries of the Hadamard product from the Kronecker product, and we show that this permutation also enables the implementation of the vectorization operator, which maps a matrix to a column vector.

In most of the previous works, the matrix $\A$ to be block-encoded is assumed to be square, with dimension a power of two. In this paper, we relax both assumptions and adopt the following generalized definition of block-encodings.
\begin{defn}\label{def:bc}
    Let matrix $\A\in \mathbb C^{M\times N}$. A unitary matrix $\U_{\A}\in \mathbb C^{2^{\scriptstyle{a}}\times 2^{\scriptstyle{a}} }$ is an $(\alpha, a)$-block-encoding of matrix $\A$ if
    \[\U_{\A} = \begin{bmatrix}
        \frac1\alpha \A&*\\ *&*
    \end{bmatrix},\]
    where $a$ denotes the total number of qubits on which $\U_{\A}$ acts, rather than the number of ancilla qubits. 
\end{defn}

\subsection{Kronecker product}\label{subsec:kron_prod}
Let $\B\in \mathbb C^{M_b\times N_b}$ and $\C\in \mathbb C^{M_c\times N_c}$. Suppose the block-encodings of $\B$ and $\C$ are, respectively, 
\[\U_{\B} = \begin{bmatrix}
    \B&*\\ *&*
\end{bmatrix}\in \mathbb C^{2^{\scriptstyle{b}} \times 2^{\scriptstyle{b}} }, \quad \U_{\C} = \begin{bmatrix}\C&*\\ *&*\end{bmatrix} \in \mathbb C^{2^{\scriptstyle{c}}\times 2^{\scriptstyle{c}} }\]
Then, the Kronecker product of $\U_{\B}$ and $\U_{\C}$ admits
\[\U_{\B}\otimes \U_{\C} = \begin{bmatrix}
    \B\otimes \begin{bmatrix}\C&*\\ *&*\end{bmatrix}& * \\
    * & * 
\end{bmatrix},
\quad \B \otimes \begin{bmatrix}\C&*\\ *&*\end{bmatrix} = \begin{bmatrix}
        b_{00} \C & * & b_{01} \C & * & \cdots \\
        * & * & * & * & \cdots \\
        b_{10} \C & * & b_{11} \C & * & \cdots\\
        * & * & * & * & \cdots \\
        \vdots& \vdots& \vdots& \vdots& \ddots
    \end{bmatrix},\]
where $b_{ij}$ denotes the $(i,j)$-entry of $\B$. Thus, while $\U_{\B}\otimes \U_{\C}$ contains all entries of $\B\otimes \C$, these entries are interleaved with undesired star blocks. 

Our target is to obtain a block-encoding whose top-left block is exactly $\B\otimes \C$, we construct permutation matrices (acting on rows and columns) that reorder the basis so as to gather the $b_{ij}\C$ blocks contiguously, i.e., find permutation unitaries $\boldsymbol \Pi_{\text{row}}$ and $\boldsymbol \Pi_{\text{col}}$ such that
\[\boldsymbol \Pi_{\text{row}}\left(\U_{\B}\otimes \U_{\C}\right)\boldsymbol \Pi_{\text{col}}^\dagger = \begin{bmatrix}
    \B\otimes \C & *\\ * & * 
\end{bmatrix},\]
where $\dagger$ denotes the Hermitian conjugate. The required permutations admit an efficient implementation on the qubit registers. To introduce the construction precisely and analyze its gate complexity, we present the following lemmas.
\begin{lem}\label{lem:Perm}
    Let $N = 2^n$ and let $\PP \in \mathbb C^{N \times N}$ be a permutation unitary such that
    \[\PP|i\rangle = |(i + d)\bmod N\rangle, \quad i\in \left\{0, 1, \dots, N-1\right\},\]
    for a fixed integer $d$. Then $\PP$ can be implemented as an addition of the classical constant $d$ modulo $N$ on $n$ qubits:
    \begin{itemize}
        \item using the ripple-carry adder of \cite{cuccaro2004newquantumripplecarryaddition} with $O(n)$ gates, $O(n)$ depth, and $n+1$ ancilla qubits; or
        \item using a QFT-based construction \cite{draper2000additionquantumcomputer,10.5555/2011517.2011525} with $O(n^2)$ gates, $O(n)$ depth, and no ancilla qubits. 
    \end{itemize}
\end{lem}
\begin{lem}\label{lem:move}
    Let 
    \[\vv_i = \begin{bmatrix}
        \aaa_i\\ *
    \end{bmatrix} \in \mathbb C^{2^{\scriptstyle{s}}}, \quad \aaa_i \in \mathbb C^M, M < 2^s, \quad \forall i = 0, \dots, 2^d - 1,\]
    and define $D = 2^d$. There exists a quantum circuit $\boldsymbol \Pi$ such that 
    \[ \boldsymbol \Pi  \begin{bmatrix}
        \vv_0\\ \vv_1\\ \vdots\\ \vv_{D-1}
    \end{bmatrix} = \begin{bmatrix}
        \aaa_0\\ \aaa_1\\\vdots\\ \aaa_{D-1}\\
        *\\\vdots\\ *
    \end{bmatrix}.\]
    The gate complexity of $\boldsymbol \Pi$ depends on $M$:
    \begin{itemize}
        \item if $M = 2^t$, $\boldsymbol \Pi$ requires $2(d - 1)$ CNOT gates;
        \item if $M > 2^{s - 1}$, $\boldsymbol \Pi$ requires $O(ds^2)$ one- or two-qubit gates;
        \item if $2^{t-1} < M < 2^t$ for some $t < s$, $\boldsymbol \Pi$ requires $O(dt^2)$ one- or two-qubit gates.
    \end{itemize}
    In summary, the gate complexity of $\boldsymbol \Pi$ is bounded by $O(d(\log M)^2)$.
\end{lem}
\begin{pf}
    For each pair $(\vv_i, \vv_{i+1})$, there exists a permutation matrix $\PP\in \mathbb R^{2^{\scriptstyle{s+}1}\times 2^{\scriptstyle{s+}1}}$ such that  
    \begin{equation}\label{eq:perm}
        \PP \begin{bmatrix}\vv_i\\ \vv_{i+1}\end{bmatrix} = \begin{bmatrix} \aaa_i\\ \aaa_{i+1}\\ * \\ * \end{bmatrix}.
    \end{equation}
    That is, $\PP$ extracts the $M$-dimensional sub-blocks $\aaa_i, \aaa_{i+1}$ from the vectors $\vv_i, \vv_{i+1}$ and moves them to the top of the new block. Applying this transformation pairwise in tensor-product form, we obtain:
    \[\I_{2^{d-2}}\otimes \PP \begin{bmatrix}
        \aaa_0\\ *\\ \aaa_1\\ *\\\vdots\\ \aaa_{D - 2} \\ * \\ \aaa_{D-1}\\ *
    \end{bmatrix} = \begin{bmatrix}
        \aaa_0\\ \aaa_1\\ *\\ *\\\vdots \\ \aaa_{D-2}\\ \aaa_{D-1}\\ *\\ *
    \end{bmatrix}\quad \text{and}\quad \I_{2^{d-3}}\otimes \PP\otimes \I_2 \begin{bmatrix}
    \aaa_0\\ \aaa_1\\ * \\ *\\\vdots \\ \aaa_{D-2}\\ \aaa_{D-1}\\ *\\ *
    \end{bmatrix} = \begin{bmatrix}
    \aaa_0\\ \aaa_1\\ \aaa_2\\ \aaa_3\\ *\\ *\\ *\\ *\\\vdots
    \end{bmatrix}. \]
    Iterating this construction for $d-1$ layers, we obtain the overall unitary
    \[ \boldsymbol \Pi = \prod_{i=0}^{d-2} \I_{2^{d-2-i}} \otimes \PP \otimes \I_{2^i},\]
    which rearranges all the $\aaa_i$ blocks consecutively at the top of the output vector, yielding the desired form. 
    
    It remains to specify the permutation $\PP$ depending on the size $M$.
    \begin{enumerate}
        \item Case $M = 2^t$: we may set $\PP = \T\otimes I_{2^{\scriptstyle{t}}}$, where $\T$ is a permutation on $s+1 -t$ qubits specified by
        \[\begin{aligned}
            \T |00\cdots 00\rangle_{s+1-t} &= |00\cdots 00\rangle_{s+1-t},\\
            \T |10\cdots 00\rangle_{s+1-t} &= |00\cdots 01\rangle_{s+1-t}. 
        \end{aligned}\]
        Since the operator $\T$ acts nontrivially only on the first and last qubits of $|\cdot \rangle_{s+1-t}$, the construction reduces to finding a two-qubit permutation $\widetilde \T$ such that
        \[\widetilde \T |00\rangle = |00\rangle , \quad \text{and} \quad \widetilde \T|10\rangle = |01\rangle.\]
        Therefore, $\widetilde \T$ must be of the form
        \[\widetilde \T = \begin{bmatrix}1&0&0&0\\ 0&0&1&0\\ 0& * & 0 & * \\ 0 & * & 0 & *\end{bmatrix}.\]
        Restricting to the permutation matrices, only two candidates exist:
        \begin{equation}\label{eq:t}
            \widetilde \T_1 = \begin{bmatrix}1&0&0&0\\ 0&0&1&0\\ 0& 1 & 0 & 0 \\ 0 & 0 & 0 & 1\end{bmatrix}, \quad \widetilde \T_2 = \begin{bmatrix}1&0&0&0\\ 0&0&1&0\\ 0& 0 & 0 & 1 \\ 0 & 1 & 0 & 0\end{bmatrix}.
        \end{equation}
        Here, $\widetilde \T_1$ is exactly the SWAP gate, requiring $3$ CNOT gates. In contrast, $\widetilde \T_2$ factorizes as 
        \begin{equation}\label{eq:T}
            \widetilde \T_2 = \begin{bmatrix}1&0&0&0\\ 0&0&1&0\\ 0&0&0&1 \\ 0&1&0&0\end{bmatrix} = \begin{bmatrix}
                1&0&0&0\\0&0&0&1\\0&0&1&0\\0&1&0&0
            \end{bmatrix}\cdot \begin{bmatrix}1&0&0&0\\0&1&0&0\\0&0&0&1\\0&0&1&0\end{bmatrix},
        \end{equation}
        which is the product of two CNOT gates. Therefore, the better choice is $\widetilde \T = \widetilde \T_2$, and consequently $\T$ (and hence $\PP$) can be implemented using $2$ CNOT gates. Since the overall circuit $\boldsymbol \Pi$ requires $(d-1)$ such permutations, the total cost is $2(d - 1)$ CNOT gates.
        \item Case $M > 2^{s-1}$: let $N = 2^s - M$. We choose $\PP$ as 
        \begin{equation}\label{eq:P}
            \begin{aligned}
            \PP = \begin{bmatrix}
            \I_M & 0 & 0& 0\\
            0 & 0 & \I_M& 0\\
            0 & 0 & 0 & \I_N\\
            0 & \I_N & 0 & 0
        \end{bmatrix} = \PP_1 \begin{bmatrix}\PP_2\\& \I_{2^s}\end{bmatrix},
        \end{aligned}
        \end{equation}
        where 
        \[\PP_1 := \begin{bmatrix}
            0 & \I_M & 0& 0\\
            0 & 0 & \I_M& 0\\
            0 & 0 & 0 & \I_N\\
            \I_N & 0 & 0 & 0
        \end{bmatrix} \in \mathbb R^{2^{\scriptstyle{s+} 1}\times 2^{\scriptstyle{s+} 1}}, \quad \PP_2 := \begin{bmatrix}
            0 & \I_N\\
            \I_M & 0
        \end{bmatrix}\in \mathbb R^{2^{\scriptstyle{s}}\times 2^{\scriptstyle{s}} }. \]
        By construction, $\PP_1$ and $\PP_2$ are cyclic shift operators:
        \[\begin{aligned}
            \PP_1 |i\rangle &= |(i - N)\bmod 2^{s+1}\rangle,\quad \forall i \in \{ 0, \dots, 2^{s+1} - 1\},\\
            \PP_2 |j\rangle &= |(j - M)\bmod 2^s\rangle, \quad \forall j \in \{ 0, \dots, 2^s - 1\}.
        \end{aligned}\]
        By \Cref{lem:Perm}, both $\PP_1$ and $\PP_2$ can be implemented using $O(s^2)$ elementary gates. Since $\PP_2$ is the top-left block of the second matrix in \eqref{eq:P}, we need the controlled version of $\PP_2$, which can also be implemented using $O(s^2)$ elementary gates. Hence $\PP$ also has gate complexity $O(s^2)$. Consequently, the full transformation $\boldsymbol \Pi$ can be implemented with gate complexity $O(ds^2)$.
        \item Case $2^{t-1} < M < 2^t$ for some $t < s$: we first embed $\aaa_i\in \mathbb C^M$ into a $2^t$-dimensional vector 
        \[\widetilde \aaa_i := \begin{bmatrix}\aaa_i \\ *\end{bmatrix} \in \mathbb C^{2^{\scriptstyle{t}}},\]
        so that $\widetilde \aaa_i$ conincides with the top $2^t$-dimensional block of $\vv_i$. Now consider two consecutive blocks $\vv_i, \vv_{i+1}$. By applying the construction from the case $M = 2^t$, we can transform the vector 
        \[\begin{bmatrix}\vv_i\\ \vv_{i+1}\end{bmatrix} \quad \mapsto \quad \begin{bmatrix}
            \widetilde \aaa_i \\ \widetilde \aaa_{i+1} \\ * \\ *
        \end{bmatrix}\]
        using $2$ CNOT gates. Next, we restrict attention to the top two blocks 
        \[\begin{bmatrix}
            \widetilde \aaa_i\\ \widetilde \aaa_{i+1}
        \end{bmatrix} \in \mathbb C^{2^{\scriptstyle{t+1}} }.\]
        By applying the construction from the case $M > 2^{s-1}$, we can transform it into the desired form using $O(t^2)$ elementary gates. Therefore, the total gate complexity of implementing $\PP$ is $O(t^2)$, and the overall gate complexity of $\boldsymbol \Pi$ is $O(dt^2)$.
    \end{enumerate}\qed
\end{pf}
The idea of \Cref{lem:move} naturally extends to block-structured matrices: by applying suitable permutations to both the rows and columns, we can align the desired submatrices into the top-left corner of a larger matrix. The following corollary formalizes this extension.
\begin{cor}\label{cor:1}
    Suppose the matrix $\V_{ij}$ has the form
    \[\V_{ij} = \begin{bmatrix}
        \A_{ij}& * \\
        * & *
    \end{bmatrix} \in \mathbb C^{2^{\scriptstyle{s}_{\scriptscriptstyle 1}}\times 2^{\scriptstyle{s}_{\scriptscriptstyle 2}}}, \quad \forall i = 0, \dots, 2^{d_1} - 1, \quad j = 0, \dots, 2^{d_2} - 1,\]
    where $\A_{ij}\in \mathbb C^{M_1\times M_2}$. Define 
    \[D_1 = 2^{d_1}, \quad D_2 = 2^{d_2}, \quad \V = \begin{bmatrix}
        \V_{11}& \cdots & \V_{1, D_2 - 1}\\
        \vdots &       & \vdots \\
        \V_{D_1-1, 1}& \cdots & \V_{D_1-1,D_2-1}
    \end{bmatrix}.\]
    Then, there exist quantum circuits $\boldsymbol \Pi_1$ and $\boldsymbol \Pi_2$ such that 
    \[\boldsymbol \Pi_1 \V \boldsymbol \Pi_2^\dagger = \begin{bmatrix}
        \A_{11}& \cdots & \A_{1, D_2 - 1}& *\\
        \vdots &       & \vdots & * \\
        \A_{D_1-1, 1}& \cdots & \A_{D_1-1,D_2-1}& *\\
        * & \cdots & * & * 
    \end{bmatrix}.\]
    In general, the gate complexity of implementing $\boldsymbol \Pi_\alpha$ is $O(d_\alpha(\log M_\alpha)^2)$ for $\alpha \in \{1, 2\}$. In the special case where $M_{\alpha}$ is a power of two, the gate complexity reduces to $O(d_\alpha)$.
\end{cor}

To obtain the block-encoding
\[\begin{bmatrix}
    \B\otimes \C & * \\
    * & * 
\end{bmatrix},\]
from $\U_{\B}\otimes \U_{\C}$, we first define 
\[d_1 = \min\{d\in \mathbb N: 2^d \ge M_b\}, \quad \text{and}\quad d_2 = \min\{d\in \mathbb N: 2^d \ge N_b\},\]
and denote the top-left $2^{d_1}\times 2^{d_2}$ block of $\U_{\B}$ as $\widetilde \B$. The desired matrix $\B\otimes \C$ is the top-left $M_bM_c \times N_bN_c$ block of
\[\begin{bmatrix}\widetilde \B \otimes \C& *\\ * & * \end{bmatrix}.\]
Notice that 
\[\widetilde b_{ij} \U_{\C} = \begin{bmatrix}\widetilde b_{ij} \C & * \\ * & * \end{bmatrix}\in \mathbb C^{2^{\scriptstyle{c}}\times 2^{\scriptstyle{c}}}, \quad \forall i = 0, \dots, 2^{d_1 - 1}, \quad j = 0, \dots, 2^{d_2} - 1,\]
where $\widetilde b_{ij}$ denotes the $(i, j)$-entry of $\widetilde \B$, $\widetilde b_{ij} \C \in \mathbb C^{M_c\times N_c}$, and thus
\[\U_{\B} \otimes \U_{\C} = \begin{bmatrix}
    \widetilde \B\otimes \U_{\C}& * \\ * & * 
\end{bmatrix}.\]
In general, applying \Cref{cor:1} yields the desired block-encoding with gate complexity 
\[O\left(\log M_b \left(\log M_c\right)^2 + \log N_b \left(\log N_c\right)^2\right),\]
which can be reduced to $O\left(\log M_b N_b\right)$ if $M_c$ and $N_c$ are powers of two.
\begin{rmk}
    The method of \cite{PhysRevA.102.052411} implements block-encodings of Kronecker products, but it is restricted to the setting where both $\B$ and $\C$ are square matrices with power-of-two dimensions. In its construction \cite{PhysRevA.102.052411}, the target blocks are moved to the top-left corner using SWAP operators, each of which requires three CNOT gates. We extend this construction to the more general case where $\B$ and $\C$ are of arbitrary sizes. When both the row and column dimensions of $\C$ are powers of two, our circuit retains the same structure as in \cite{PhysRevA.102.052411} but substitutes the SWAP operator with the two-CNOT operator $\widetilde \T_2$ from \cref{eq:T}, thereby reducing the two-qubit gate count.
\end{rmk}

\subsection{Matrix-matrix product}
\begin{figure}
    \centering
    \[\Qcircuit @C=1em @R=1em{
        \lstick{|0\rangle_{b-s}}& \qw & \qw & \sgate{\U_{\B}}{2} & \qw\\
        \lstick{|0\rangle_{c-s}}& \multigate{1}{\U_{\C}} & \qw & \qw & \qw\\
        \lstick{|\psi\rangle_s}& \ghost{\U_{\C}} & \qw & \gate{\U_{\B}} & \qw} 
    \]
    \caption{Quantum circuit for standard implementation of matrix-matrix product.}
    \label{fig:mm_prod}
\end{figure}
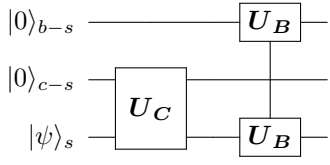

Let $\B\in \mathbb C^{M\times K}$, $\C\in \mathbb C^{K\times N}$, and suppose we have block-encodings
\[\U_{\B} = \begin{bmatrix}
    \B&*\\ *&*
\end{bmatrix}\in \mathbb C^{2^{\scriptstyle{b}} \times 2^{\scriptstyle{b}} }, \quad \U_{\C} = \begin{bmatrix}\C&*\\ *&*\end{bmatrix} \in \mathbb C^{2^{\scriptstyle{c}}\times 2^{\scriptstyle{c}} }.\]
Our goal is to construct a block-encoding of the product $\B\C$. Even if $b = c$, a direct multiplication does not suffice, since
\[\U_{\B} \U_{\C} = \begin{bmatrix}\B\C + * & *\\ * & *\end{bmatrix},\]
so the top-left block contains undesired terms. In the special case where $M = K = N = 2^s$, the standard construction \cite{10.1145/3313276.3316366} empolys the circuit shown in \Cref{fig:mm_prod}. 
This construction can be understood in terms of matrix representation: additional ancilla qubits enforce zero-padding in $\U_{\B}$ and $\U_{\C}$ so that the unwanted terms $*$ are eliminated. However, this procedure introduces more zero-padding than necessary, thereby consuming extra qubits and leading to inefficiency. 

In the following proposition, we present a qubit-efficient method to construct block-encodings of matrix products. Our approach applies to general rectangular compatiable matrices $\B$ and $\C$. 
\begin{prop}
    Let $\B\in \mathbb C^{M\times K}$, $\C\in \mathbb C^{K\times N}$, and suppose we are given block-encodings
    \[\U_{\B} = \begin{bmatrix}
        \B&*\\ *&*
    \end{bmatrix}\in \mathbb C^{2^{\scriptstyle{b}}\times 2^{\scriptstyle{b}} }, \quad \U_{\C} = \begin{bmatrix}\C&*\\ *&*\end{bmatrix} \in \mathbb C^{2^{\scriptstyle{c}} \times 2^{\scriptstyle{c}} }.\]
    Then there exists a quantum circuit that implements the block-encoding of $\B\C$ using $\max\{b, c\} + 1$ qubits, with an additional gate complexity of $O(\min\{b^2, c^2\})$ beyond that required for implementing $\U_{\B}$ and $\U_{\C}$
\end{prop}
\begin{pf}
    We begin with two simple cases:
    \begin{itemize}
        \item If $K = 2^b$, then $(\I_{2^{c-b}}\otimes \U_{\B})\U_{\C}$ block-encodes $\B\C$.
        \item If $K = 2^c$, then $\U_{\B}\left(\I_{2^{b - c}}\otimes \U_{\C}\right)$ block-encodes $\B\C$.
    \end{itemize}
    In both cases, the block-encoding of $\B\C$ requires only $\max\{b,c\}$ qubits, with no additional gate complexity beyond that of $\U_{\B}$ and $\U_{\C}$. 
    
    In the nontrivial case $K < \min\{2^b, 2^c\}$, we assume without loss of generality that $b \ge c$. (If $b < c$, one may instead construct a block-encoding of $\C^\dagger \B^\dagger$ and then invert the circuit.) We define the auxiliary block-encoding 
    \[\U_{\widetilde \C} = \begin{bmatrix}\widetilde \C & * \\ * & * \end{bmatrix}\in \mathbb C^{2^{\scriptstyle{t}} \times 2^{\scriptstyle{t}}}, \quad \text{where} \quad \widetilde \C = \begin{bmatrix}\C\\ 0\end{bmatrix} \in \mathbb C^{2^{\scriptstyle{b}} \times N}, \quad t = \max\{b, c\} + 1 = b + 1.\]
    Then,
    \[\left(\I_{2}\otimes\U_{\B}\right)\U_{\widetilde\C}=\begin{bmatrix}\B\C& * \\ * & * \end{bmatrix},\]
    is the desired block-encoding of the product $\B\C$. Note that the resulting block-encoding costs only one more qubit than $\U_{\B}$. It remains to show how to construct the block-encoding $\U_{\widetilde \C}$. We begin with the Kronecker product $\I_{2^{t-c}}\otimes \U_{\C}$, whose first $N$ columns take the form
    \[ \begin{bmatrix}
        \C\\ * \\ 0 \\ 0 \\\vdots \\ 0 \\ 0
    \end{bmatrix} = |0^{t-c}\rangle \otimes \begin{bmatrix}\C\\ *\end{bmatrix} =: \V.\]
    Our goal is to rearrange $\V$ into a block structure where $\C$ is isolated at the top, separated by at least $2^{b} - K$ rows of zeros from the $*$ terms. More precisely, we seek a $t$-qubit permutation $\PP$ such that 
    \begin{equation}\label{eq:pv}
        \PP \V = |0\rangle \otimes |0^{t-c-1}\rangle \otimes \begin{bmatrix}\C \\ 0\end{bmatrix} + |1\rangle \otimes |0^{t-c-1}\rangle \otimes \begin{bmatrix}0 \\ *\end{bmatrix}.
    \end{equation}
    Notice that only the first qubit and the last $c$ qubits are involved in this transformation; the intermediate $t-c-1$ qubits remain untouched. Therefore, it suffices to define a permutation acting on these $(c+1)$ qubits. In the computational basis of these qubits, consider
    \[\widetilde \PP := \begin{bmatrix}
        \I_K & 0 & 0 & 0\\
        0 & 0 & 0 & \I_L\\
        0 & 0 & \I_K & 0\\
        0 & \I_L & 0 & 0 
    \end{bmatrix} \in \mathbb R^{2^{\scriptstyle{c+1}}\times 2^{\scriptstyle{c+1}} },\quad L = 2^{c+1} - K.\]
    This permutation achieves the transformation
    \[|0\rangle \otimes \begin{bmatrix}\C\\ *\end{bmatrix}\quad \stackrel{\widetilde {\PP}}{\longrightarrow} \quad |0\rangle \otimes \begin{bmatrix}\C \\ 0\end{bmatrix} + |1\rangle \otimes \begin{bmatrix}0 \\ *\end{bmatrix},\]
    which matches exactly the desired block structure in \cref{eq:pv}. For implementation, $\widetilde \PP$ can be factorized as  
    \[\widetilde \PP = \begin{bmatrix}0 & 0 & 0 & \I_{K}\\ \I_L & 0 & 0 & 0 \\ 0 & \I_K & 0 & 0 \\ 0 & 0 & \I_L & 0\end{bmatrix}  \left(\begin{bmatrix}0&1\\1&0\end{bmatrix} \otimes \begin{bmatrix}0 & \I_L\\ \I_K & 0\end{bmatrix}\right).\]
    Each factor can be implemented using \Cref{lem:Perm} with gate complexity $O(c^2)$. Since $t-1 = b$, it follows that $\PP (\I_{2^{t-c}}\otimes \U_{\C})$ is a block-encoding of $\U_{\widetilde \C}$, and thus
    \[\left(\I_{2}\otimes \U_{\B}\right) \PP \left(\I_{2^{t - c}}\otimes \U_{\C}\right)\]
    provides a block-encoding of $\B\C$. In summary, this block-encoding requires $\max\{b,c\} + 1$ qubits and can be implemented with $O\left(\min\{b^2, c^2\}\right)$ gate complexity.\qed
\end{pf}

\begin{ex}\label{ex:1}
    Suppose $M = K = N = 2^s$. In this case, we can choose
    \[\widetilde \PP = \begin{bmatrix}1&0&0&0\\ 0&0&0&\I_{2^{c-s} - 1}\\ 0&0&1&0\\ 0&\I_{2^{c-s}-1}&0&0\end{bmatrix}\otimes \I_{2^s},\]
    which can be implemented using a single NOT gate and a multi-controlled NOT gate. The quantum circuit for the block-encoding of $\B\C$ is then given in \Cref{fig:qubit_saving_mm}.
    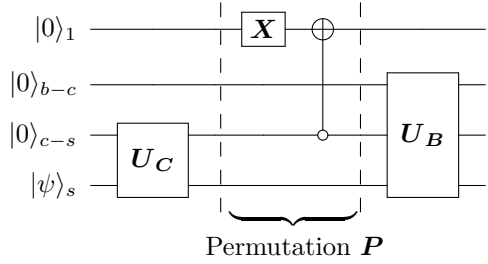
\begin{figure}
        \centering
         \[\Qcircuit @C=1em @R=1em{
            \lstick{|0\rangle_1}& \qw & \qw \barrier[-1.6em]{3}& \gate{\X} & \targ \barrier[0em]{3} &  \qw & \qw & \qw\\
            \lstick{|0\rangle_{b-c}}& \qw & \qw & \qw & \qw & \qw & \multigate{2}{\U_{\B}} & \qw\\
            \lstick{|0\rangle_{c-s}}& \multigate{1}{\U_{\C}} & \qw & \qw & \ctrlo{-2} & \qw & \ghost{\U_{\B}} & \qw\\
            \lstick{|\psi\rangle_s}& \ghost{\U_{\C}}& \qw &\qw &\qw & \qw & \ghost{\U_{\B}} & \qw \gategroup{4}{4}{4}{5}{2.5em}{_\}} \\
            &&&\dstick{\quad \quad \text{ Permutation }\PP}} 
         \]
        \caption{Quantum circuit for qubit-efficient implementation of matrix-matrix product. The subscript of each ket indicates the number of qubits in that ket.}
        \label{fig:qubit_saving_mm}
    \end{figure}
    To verify correctness, consider its action on the input state $|0\rangle_{b-s+1}|\psi\rangle_s$:
    \[\begin{aligned}
        |0\rangle_1 |0\rangle_{b-c} |0\rangle_{c-s} |\psi\rangle_s & \longrightarrow |1\rangle_1 |0\rangle_{b-c} \left(|0\rangle_{c-s} \C|\psi\rangle_s + |g \rangle_c \right)\\
        & \longrightarrow |0\rangle_1 |0\rangle_{b-c} |0\rangle_{c-s} \C|\psi\rangle_s + |1\rangle_1 |0\rangle_{b-c} |g\rangle_c\\
        & \longrightarrow |0\rangle_1 |0\rangle_{b-c} |0\rangle_{c-s} \B\C|\psi\rangle_s + |0\rangle_1 |g'\rangle_b + |1\rangle_1 |g''\rangle_b,
    \end{aligned}\]
    where $|g\rangle_c$ is orthogonal to $|0\rangle_{c-s}\otimes I_{2^s}$, and $|g'\rangle_b$ is orthogonal to $|0\rangle_{b-s}\otimes I_{2^s}$. 
\end{ex}
\begin{rmk}
    The existing method for implementing matrix-matrix product of block-encoded matrices \cite{10.1145/3313276.3316366} applies only to the case where both matrices are square of the same size, and this dimension must be a power of two. In contrast, our method generalizes to arbitrary matrix sizes without such restrictions. Below, we provide a comparison of the required number of qubits and the gate complexity between the two approaches.

    Suppose $M = K = N = 2^s$ are powers of two. The existing method requires $b + c - s$ qubits, while our method requires
    \[t = \max\{b, c\} + 1\]
    qubits. Since $2^s = K < 2^{\min\{b, c\}}$, we have 
    \[b + c - s > b + c - \min\{b, c\} = \max\{b, c\},\]
    and thus 
    \[b + c - s \ge \max\{b, c\} + 1.\] 
    Therefore, our method always uses fewer qubits. Regarding the gate complexity, the existing method does not incur additional costs beyond the block-encodings $\U_{\B}$ and $\U_{\C}$. In contrast, our method requires $O(\min\{b^2, c^2\})$ elementary quantum gates in general.
\end{rmk}
\begin{rmk}
    A qubit-efficient implementation of the Kronecker product, inspired by our matrix-matrix product construction, is described in \ref{app:A}. 
\end{rmk}
\subsubsection{Product of a sequence of matrices}
The qubit-efficient advantage of our method becomes more pronounced when considering the product of a sequence of matrices. Suppose we are given matrices $\A_1, \dots, \A_n\in \mathbb C^{2^{\scriptstyle{s}}\times 2^{\scriptstyle{s}}}$, and their block-encodings have sizes $2^{a_1}\times 2^{a_1}, \dots, 2^{a_n}\times 2^{a_n}$, respectively. Without loss of generality, assume $s < a_i$ for all $i$. Our goal is to construct a block-encoding of the product $\A_1 \A_2\cdots  \A_n$. 

If we apply the existing method \cite{10.1145/3313276.3316366} iteratively, the required number of qubits is $\sum_{i = 1}^n a_i - (n - 1)s$. Since $s < a_i$ for all $i$, this number scales as $\max_{1\le i\le n} a_i + O(n-1)$. In contrast, our method can reduce the additive overhead from $O(n - 1)$ to $\lceil \log_2 n\rceil $. This improvement relies on choosing an optimal multiplication order via dynamic programming. Define $m_{ij}$ as the minimum number of qubits required to block-encode the product $\A_i\A_{i+1}\cdots \A_j$. Then $m_{ij}$ satisfies the recurrence 
\begin{equation}
    m_{ij} = \min_{i \le r < j} \left\{ \max \left\{m_{ir}, m_{r+1, j}\right\} + 1\right\}.
\end{equation}
We now show by induction that 
\begin{equation}\label{eq:induction}
    m_{ij} \le \max_{i\le r\le j} a_{r} + \left\lceil \log_2(j - i + 1)\right \rceil.
\end{equation}
For $j = i + 1$,
\[m_{i, i+1} = \max \left\{a_i, a_{i+1}\right\} + 1,\]
which clearly satisfies \cref{eq:induction}. Suppose the bound holds for all intervals with length samller than $k$. Then for $j = i + k$ and $i\le r < j$, we have 
\[\begin{aligned}
    m_{ij} &\le \max \left\{m_{ir}, m_{r+1, j}\right\} + 1\\
    &\le \max \left\{\max_{i\le l\le r} a_{l} + \left\lceil \log_2(r - i + 1)\right \rceil, \max_{r+1 \le l\le j} a_{l} + \left\lceil \log_2(j - r)\right \rceil \right\} + 1 \\
    &\le \max_{i\le l \le j} a_l + \max\left\{\left \lceil \log_2(r-i+1)\right \rceil , \left \lceil \log_2(j-r)\right \rceil \right\} + 1.
\end{aligned}\]
Note that there exists an index $r^*$ such that the two subintervals $[i, r^*]$ and $[r^* + 1, j]$ are approximately balanced in length. In particular, we can choose $r^*$ so that 
\[\max\{r^*-i+1, j-r^*\} \le \left\lceil \frac{j-i+1}{2}\right\rceil.\]
For the choice of $r^*$, we obtain
\[\max\left\{\left \lceil \log_2(r^*-i+1)\right \rceil , \left \lceil \log_2(j-r^*)\right \rceil \right\} + 1 \le \left \lceil \log_2 \left(\left\lceil \frac{j-i+1}{2}\right\rceil\right) + 1\right \rceil.\]
Now we analyze the right-hand side depending on the parity of $j - i + 1$:
\begin{itemize}
    \item If $j-i+1$ is even, then 
    \[ \left \lceil \log_2 \left\lceil \frac{j-i+1}{2}\right\rceil + 1\right \rceil =  \left \lceil \log_2 (j-i+1)\right \rceil.\]
    \item If $j-i+1$ is odd, then 
    \[\left \lceil \log_2 \left\lceil \frac{j-i+1}{2}\right\rceil + 1\right \rceil =  \left \lceil \log_2 (j-i+2)\right \rceil = \left \lceil \log_2 (j-i+1)\right \rceil,\]
    where the second equality holds due to $j - i + 1$ is odd and $j > i + 1$. 
\end{itemize}
Therefore, in both cases, we have 
\[\max\left\{\left \lceil \log_2(r^*-i+1)\right \rceil , \left \lceil \log_2(j-r^*)\right \rceil \right\} + 1 \le \left \lceil \log_2 (j-i+1)\right \rceil.\]
Substitute this bound into the recurrence gives
\[m_{ij} \le \max_{i\le l\le j} a_l + \left\lceil \log_2 (j-i+1) \right \rceil, \quad \forall j > i+1.\]
Together with the base case $j = i + 1$, this complete the induction for \cref{eq:induction}. In particular, for the full product we obtain 
\[m_{1n} \le \max_{1\le r\le n} a_r + \left\lceil \log_2 n\right \rceil.\]
Moreover, a simple upper bound to the additional gate complexity is 
\[O\left(n \left(\log n + \max_{i}a_i - a\right)^2\right).\]
Since the baseline cost of implementing $n$ block-encodings is $\Omega(n)$, this additional overhead grows only polylogarithmically and is therefore acceptable. 

\subsection{Hadamard product}
In this section, we extend the implementation of the Hadamard product proposed in \cite{guo2024quantumlinearalgebraneed} to matrices of arbitrary dimensions. In addition, we present implementations for convolution operation and vectorization operator, which are closely related to the construction of the Hadamard product. 

Let $\B, \C \in \mathbb C^{M\times N}$ be two matrices. Suppose the block-encodings of $\B$ and $\C$ are, respectively,
\[\U_{\B} = \begin{bmatrix}
    \B&*\\ *&*
\end{bmatrix}\in \mathbb C^{2^{\scriptstyle{b}}\times 2^{\scriptstyle{b}} }, \quad \U_{\C} = \begin{bmatrix}\C&*\\ *&*\end{bmatrix} \in \mathbb C^{2^{\scriptstyle{c}} \times 2^{\scriptstyle{c}} }\]
Without loss of generality, we assume $M = 2^l$ and $N = 2^r$, since the block-encoding of the Hadamard product for $\B$ and $\C$ can be obtained from that of the corresponding top-left submatrices of $\U_{\B}$ and $\U_{\C}$ that contain $\B$ and $\C$. Let $\circ$ denote the Hadamard product, we have 
\[\begin{aligned}
    \B\circ \C &= \sum_{i=0}^{M-1} \sum_{j = 0}^{N - 1} b_{ij} c_{ij} |i\rangle_l \langle j|_r\\
    &= \sum_{i=0}^{M - 1} \sum_{j = 0}^{N - 1}  |i\rangle_l \left(\langle i|_b \U_{\B} |j\rangle_b \langle i|_c \U_{\C} |j\rangle_c \right)\langle j|_r\\
    &= \sum_{i=0}^{M - 1} \sum_{j = 0}^{N - 1} |i\rangle_l \left(\langle i|_b \otimes \langle i|_c\right) \left(\U_{\B} \otimes \U_{\C}\right)\left(|j\rangle_b \otimes |j\rangle_c \right)\langle j|_r\\
    &= \left(\sum_{i = 0}^{M - 1}|i\rangle_l \left( \langle i|_b\otimes \langle i|_c\right)\right) (\U_{\B}\otimes \U_{\C}) \left(\sum_{j = 0}^{N - 1} \left(|j\rangle_b\otimes |j\rangle_c \right)\langle j|_r\right),
\end{aligned}\]
where the subscript denotes the number of qubits in that ket (or bra). Define
\begin{equation}\label{eq:T_r}
    \T_r := \sum_{j = 0}^{N - 1} \left(|j\rangle_b\otimes |j\rangle_c \right)\langle j|_r.
\end{equation}
To encode $\T_r$ into quantum circuit, let $\PP_r$ satisfy
\[\PP_r |0\rangle_b \otimes |j\rangle_c = |j\rangle_b \otimes |j\rangle_c,\quad \forall j = 0, \dots, 2^r - 1.\]
The operator $\PP_r$ replicates the bit string stored in the register $|\cdot\rangle_c$ into the register $|\cdot \rangle_b$. In general, this operation requires $c$ CNOT gates. However, since we only need to handle the bit strings smaller than $2^r -1$, it suffices to use $r$ CNOT gates. Then, we obtain
\[\begin{aligned}
    \PP_r \left(|0\rangle_{b+c-r} \otimes I_{2^r}\right) &= \sum_{j=0}^{2^r - 1} \PP_r \left(|0\rangle_{b+c-r} \otimes |j\rangle_r \langle j|_r\right) = \sum_{j = 0}^{2^r - 1} \PP_r \left(|0\rangle_{b+c-r} \otimes |j\rangle_r\right) \left(1\otimes \langle j|_r \right)\\
    &= \sum_{j = 0}^{2^r - 1} \PP_r \left(|0\rangle_{b} \otimes |j\rangle_c\right) \langle j|_r =  \sum_{j = 0}^{2^r - 1} \left(|j\rangle_{b} \otimes |j\rangle_c\right) \langle j|_r,
\end{aligned}\]
where we combine the last $c-r$ qubits in $|0\rangle_{b+c-r}$ with $|j\rangle_r$ to get $|j\rangle_c$ in the third equality. Since $N = 2^r$, the first $N$ columns of $\PP_r$ coincide with $\T_r$. Similarly, the operator
\[\T_l := \sum_{i = 0}^{M - 1} \left(|i\rangle_b\otimes |i\rangle_c \right)\langle i|_l\]
can be encoded as the first $M$ columns of a permutation matrix $\PP_l$, which can be implemented using $l$ CNOT gates. Finally,
\[\begin{aligned}
    \PP_l^\dagger \left(\U_{\B}\otimes \U_{\C}\right) \PP_r &= \begin{bmatrix}
        \T_l^\dagger \\ * 
    \end{bmatrix}\left(\U_{\B}\otimes \U_{\C}\right) \begin{bmatrix}\T_r & * \end{bmatrix}\\ 
    &= \begin{bmatrix}\T_l^\dagger \left(\U_{\B}\otimes \U_{\C}\right)\T_r &*\\ *&*\end{bmatrix} = \begin{bmatrix}
    \B\circ \C &*\\ *&*
    \end{bmatrix}.
\end{aligned}\]
Hence, the Hadamard product of two block-encodings can be realized using $l+r = O(\log (MN))$ CNOT gates.
\subsubsection{Convolution}
Let 
\[|\psi\rangle = \sum_{i = 0}^{2^n - 1} \psi_i |i\rangle , \quad |\varphi\rangle = \sum_{i = 0}^{2^n - 1} \varphi_i |i\rangle.\]
Our goal is to prepare the quantum state encoding their discrete convolution,
\[|\psi * \varphi \rangle = \frac1c\sum_{i = 0}^{2^n - 1} (\psi * \varphi)_i |i\rangle,\]
where $c$ is a normalization factor. Using the standard relation between convolution and Fourier transform $\mathcal F$, we have 
\[\psi * \varphi = \mathcal F^{-1}\left(\mathcal F(\psi) \circ \mathcal F(\varphi)\right),\]
where $\circ$ denotes the element-eise (Hadamard) product. This relation naturally motivates a quantum implementation: one first applies the Quantum Fourier Transform (QFT) to both input states, performs the Hadamard product in the Fourier domain, and finally applies the inverse QFT to obtain the convolution.

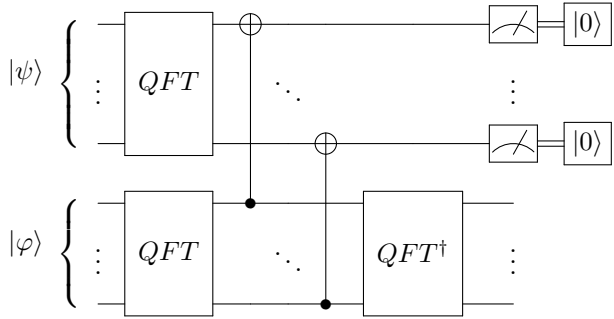
\begin{figure}
    \centering
    \[\Qcircuit @C=1em @R=1em {
        \lstick{}& &\multigate{2}{QFT} & \targ & \qw & \qw & \qw  & \meter & \cgate{|0\rangle}\\
        \lstick{}& \vdots  &\nghost{QFT} & & \ddots & && \vdots \\
        \lstick{}& &\ghost{QFT}  & \qw & \qw & \targ & \qw & \meter & \cgate{|0\rangle} \inputgroupv{1}{3}{.2em}{1.8em}{|\psi \rangle}\\
        \lstick{}& &\multigate{2}{QFT}  & \ctrl{-3} & \qw & \qw & \multigate{2}{QFT^{\dagger}} &\qw\\
        \lstick{}& \vdots &\nghost{QFT}& & \ddots & & \nghost{QFT^{\dagger}} & \vdots \\
        \lstick{}& &\ghost{QFT} & \qw & \qw & \ctrl{-3} & \ghost{QFT^{\dagger}} & \qw \inputgroupv{4}{6}{.2em}{1.5em}{|\varphi \rangle}
    }\]
    \caption{Quantum circuit of convolution. The Hadamard product between two quantum states appears as a special case of the Hadamard product between block-encodings. Here, $n$ CNOT gates implement $\PP_l$, while $\PP_r$ reduces to the identity since a quantum state has row dimension one.}
    \label{fig:convolution}
\end{figure}

The corresponding quantum circuit is shown in \Cref{fig:convolution}. The two registers are initialized in $|\psi\rangle$ and $|\varphi\rangle$. Applying the QFT in parallel transforms both registers into the Fourier basis. At this stage, a sequence of $n$ CNOT gates performs the Hadamard product between the transformed amplitudes. Finally, the inverse QFT is applied to the second register, thereby producing the quantum state corresponding to the convolution in the computational basis.
Upon measuring the first register and obtaining $|0\rangle$, the second register collapses to the state $|\psi * \varphi\rangle$. The overall cost is dominated by the QFT and its inverse, each of which requires $O(n^2)$ one- and two-qubit gates in the standard implementation. Therefore, the gate complexity of the convolution procedure is $O(n^2)$.

In summary, this construction shows how the well-known convolution-Fourier duality can be translated into the quantum setting, leveraging the Hadamard product between block-encodings together with efficient implementations of the QFT.
\subsubsection{Vectorization}\label{subsubsec:vec}
Given a matrix 
\[\A = \begin{bmatrix}
    \aaa_0 & \aaa_1 & \cdots & \aaa_{N-1}
\end{bmatrix}\in \mathbb C^{M\times N},\]
the operator $\texttt{vec}(\cdot)$ stacks the columns of $\A$ into a single long vector:
\[\texttt{vec}(\A) := \begin{bmatrix}\aaa_0\\ \aaa_1\\ \vdots \\ \aaa_{N-1}\end{bmatrix} \in \mathbb C^{MN}.\]
Our goal is to construct a block-encoding of $\texttt{vec}(\A)$ given a block-encoding of $\A$. 

\begin{figure}
    \centering
    \[\Qcircuit @C=1em @R=1em {
        \lstick{}& & \qw & \targ & \qw & \qw & \qw  & \multigate{6}{\Pi}& \qw\\
        \lstick{}& & & & \ddots & & & &  \\
        \lstick{}& &\qw  & \qw & \qw & \targ & \qw & \ghost{\Pi} & \qw \inputgroupv{1}{3}{.2em}{1.2em}{|\cdot \rangle_r}\\
         \lstick{}& & \qw  & \qw & \qw & \qw & \multigate{3}{\U_{\A}} &\ghost{\Pi} &\qw\\
        \lstick{}& & \multigate{2}{\HH^{\otimes r}} & \ctrl{-4} & \qw & \qw & \ghost{\U_{\A}} &\ghost{\Pi} &\qw \inputgroupv{4}{7}{.2em}{2.6em}{|\cdot \rangle_a}\\
        \lstick{}& \vdots & \nghost{\HH^{\otimes r}} & & \ddots & & \nghost{\U_{\A}} & & \vdots \\
        \lstick{}& & \ghost{\HH^{\otimes r}} & \qw & \qw & \ctrl{-4} & \ghost{\U_{\A}} & \ghost{\Pi} & \qw 
    }\]
    \caption{Quantum circuit of vectorization. The Hadamard layer prepares the uniform superposition, the CNOT chain realizes $\T_r$, and the permutation $\Pi$ (from \Cref{lem:move}) gathers the $\aaa_i$ blocks contiguously, resulting in a block-encoding of $\texttt{vec}(\A)$.}
    \label{fig:vectorization}
\end{figure}
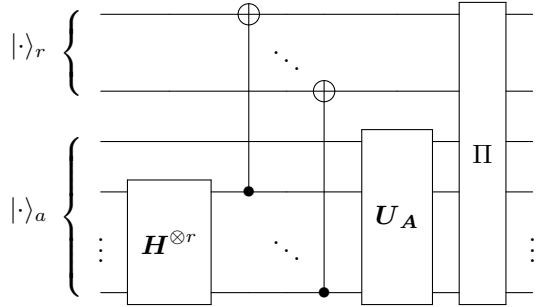

Assume first that $N = 2^r$, and let $\U_{\A}\in \mathbb C^{2^{\scriptstyle{a}}\times 2^{\scriptstyle{a}}}$ be a block-encoding of $\A$. Consider the stacked action of the first $N$ columns of $\U_{\A}$,
\[\begin{aligned}
    \begin{bmatrix}\U_{\A}|0\rangle_a \\ \U_{\A}|1\rangle_a \\ \vdots \\ \U_{\A}|N-1\rangle_a \end{bmatrix} = \left(\I_N \otimes \U_{\A}\right)\begin{bmatrix}|0\rangle_a \\ & |1\rangle_a \\ && \ddots \\ &&& |N-1\rangle_a \end{bmatrix} \begin{bmatrix}1\\ 1\\ \vdots \\ 1\end{bmatrix}.
\end{aligned}\]
The block-diagonal operator in the middle can be expressed as
\[\begin{bmatrix}|0\rangle_a \\ & |1\rangle_a \\ && \ddots \\ &&& |N-1\rangle_a \end{bmatrix} = \sum_{j = 0}^{N-1} (|j\rangle_r \langle j|_r) \otimes |j\rangle_a = \sum_{j = 0}^{N-1}\left(|j\rangle_r\otimes |j\rangle_a\right) \langle j |_r,\]
which has the same form as $\T_r$ introduced in \cref{eq:T_r}. This observation allows us to block-encode it into a simple permutation using $r$ CNOT gates. The all-ones vector of length $N$ appearing on the right-hand side is proportional to the first column of $\HH^{\otimes r}$, where $\HH$ is the Hadamard gate. Hence, applying $\HH^{\otimes r}$ to $|0\rangle_r$ prepares the required uniform superposition. The resulting stacked vector takes the form
\[\begin{bmatrix}\U_{\A}|0\rangle_a \\ \U_{\A}|1\rangle_a \\ \vdots \\ \U_{\A}|N-1\rangle_a \end{bmatrix} = \begin{bmatrix}\aaa_0 \\ * \\ \aaa_1 \\ *\\  \vdots \\ \aaa_{N-1}\\ *\end{bmatrix},\]
where $*$ represents additional ancillary components. By applying the permutation described in \Cref{lem:move}, these unwanted components can be rearranged so that the $\aaa_i$ blocks appear contiguously, giving precisely the block-encoding of $\texttt{vec}(\A)$. The full construction is illustrated in \Cref{fig:vectorization}. 

For the case where $N$ is not a power of two, one can take the smallest $r$ such that $2^r > N$, and thus the same procedure yields the block-encoding of $\texttt{vec}(\A)$. The additional gate complexity beyond that of $\U_{\A}$ is dominated by the final permutation $\Pi$, which, by \Cref{lem:move}, can be performed with gate complexity
\[O\left(r(\log M)^2\right) = O\left(\log N(\log M)^2\right).\]
If $M$ happens to be a power of two, the bound improves to $O(\log N)$.
\section{Applications}\label{sec:3}
Matrix products---including the matrix-matrix multiplication, the Kronecker product, and the Hadamard product---are fundamental operations in linear algebra and arises in a variety of contexts. Block-encoding, in turn, is a key primitive in many quantum algorithms, such as Hamiltonian simulation, quantum singular value transformation (QSVT), quantum linear system algorithms (QLSAs), and quantum walks. Naturally, matrix products between block-encodings also occur frequently in quantum computing. In this section, we present several applications of such products.
\subsection{Time-dependent Hamiltonian Simulation}
In \cite{low2019hamiltoniansimulationinteractionpicture}, a low-space-overhead simulation algorithm based on the truncated Dyson series is proposed for time-dependent quantum dynamics. To reduce space requirements, a compression gadget is introduced.
\begin{lem}[Lemma 13 in \cite{low2019hamiltoniansimulationinteractionpicture}]\label{lem:lem13}
    Let $\{\U_k: k\in [K]\}$ be a set of $K$ unitaries that encode matrices $\HH_k \in \mathbb C^{2^{\scriptstyle{n}_{\scriptscriptstyle{s}}} \times 2^{\scriptstyle{n}_{\scriptscriptstyle{s}}} }$ such that 
    \[ \left(\langle 0|_a \otimes \I_s\right) \U_k \left(|0\rangle_a \otimes \I_s\right) = \HH_k, \quad \|\HH_k\|\le 1, \quad |0\rangle_a \in \mathbb C^{2^{\scriptstyle{n}_{\scriptscriptstyle{a}}} }.\]
    Then there exists a quantum circuit $\V$ such that on input states spanned by $\{|k\rangle_b: k\in \{0, \dots, K\}\}$, 
    \[\left(\langle 0|_{ac} \otimes \I_s\right) \V \left(|0\rangle_{ac} \otimes \I_s\right) = |0\rangle \langle 0|_b \otimes \I_s + \sum_{k = 1}^K |k\rangle\langle k|_b \otimes \left(\prod_{j=1}^k \HH_j\right), \quad |k\rangle_b \in \mathbb C^{2^{\scriptstyle{n}_{\scriptscriptstyle{b}}} }, \quad |0\rangle_c \in \mathbb C^{2^{\scriptstyle{n}_{\scriptscriptstyle{c}}} },\]
    where the number of qubits $n_b \in O(n_c) = O(\log K)$ and 
    \[\prod_{j=1}^k \HH_j := \HH_k \HH_{k-1} \cdots \HH_2\HH_1.\]
    The cost of $\V$ is one query to each  controlled-controlled-$\U_k$, and $O(K(n_a + \log K))$ additional primitive quantum gates. 
\end{lem}
In the truncated Dyson series approach for time-dependent Hamiltonians, one often needs to implement sequential products of many Hamiltonian terms, controlled on an index register that select which terms appear in the series. \Cref{lem:lem13} from \cite{low2019hamiltoniansimulationinteractionpicture} guarantees that these products can be implemented with exponentially fewer ancillary qubits than a na\"ive implementation, while keeping gate complexity linear in the number of terms. Following the qubit-efficient approach for implementing matrix-matrix products, we provide a simpler construction of $\V$ that only uses controlled-$\U_k$. The asymptotic gate complexity remains the same as in the original method, but redundant controls are eliminated, yielding a more concise quantum circuit. Here is our construction.

Let $\T_{p, q}$, with $1\le p \le q\le K$, denote a quantum circuit that performs the transformation:
\[\begin{aligned}
    \T_{p, q} |0\rangle_c |k\rangle_b |0\rangle_a |\psi\rangle_s = |0\rangle_c |k - q + p - 1\rangle_b |0\rangle_a \left(\prod_{j = p}^{\min\{q, p + k - 1\}}\HH_j\right) |\psi\rangle_s + | g \rangle,\quad \forall 1-p \le k\le K,
\end{aligned}\]
where the arithmetic in $|\cdot \rangle_b$ is modulo $2^{n_b}$, and $|g\rangle$ is orthogonal to the subspace $|0\rangle_c \otimes \I_b \otimes |0\rangle_a \otimes \I_s$. To be clarified, in \Cref{tab:B2I}, we list the correspondence between bit strings and integers in the register $|\cdot\rangle_b$. 
\begin{table}[t]
    \centering
    \begin{tabular}{cccccccc}
        \toprule
        Bit string & $10\cdots 01$ & $\cdots$ & $11 \cdots 11$ & $00 \cdots 00$ & $00 \cdots 01$ & $\cdots$ & $10 \cdots 00$\\
        \midrule
        Integer &$-2^{n_b - 1} + 1$ & $\cdots$ & $-1$& $0$& $1$ & $\cdots$ & $2^{n_b-1}$\\
        \bottomrule
    \end{tabular}
    \caption{The correspondence between bit strings and integers.}\label{tab:B2I}
\end{table}
Then, if $n_b \ge 1 + \log_2 K$, we have 
\[\T_{1, K} |0\rangle_c |k\rangle_b |0\rangle_a |\psi\rangle_s = |0\rangle_c |k - K\rangle_b |0\rangle_a \left(\prod_{j = 1}^{k} \HH_j\right) |\psi\rangle_s + | g \rangle,\quad \forall 0 \le k\le K,\]
and thus the quantum circuit $\V$ can be chosen as
\[ \text{ADD}_b^{K} \T_{1, K},\]
where $\text{ADD}_b^{K}$ adds the integer $K$ to the register $|\cdot \rangle_b$. The construction of $\T_{1, K}$ proceeds in three steps:
\begin{enumerate}
    \item \textbf{Base case:} $\T_{p, p}$ can be implemented using the circuit shown in \Cref{fig:Tpp}. The gate $\text{ADD}_b^\dagger$ acts as a decrement (minus-one) operator on the register $|\cdot \rangle_b$, and can be implemented in two ways according to \Cref{lem:Perm}. For consistency with the analysis in \cite{low2019hamiltoniansimulationinteractionpicture}, we assume that $\text{ADD}_b^\dagger$ is implemented using $O(n_b)$ gates and $n_b + 1$ ancilla qubits. The operator $\U_p$ is controlled by the most significant qubit in $|\cdot \rangle_b$. Specifically, if $k \le 0$, the most significant qubit of $|k - 1\rangle_b$ is $1$ and thus the output is 
    \[|k-1\rangle_b |0\rangle_a |\psi\rangle_s.\]
    For $1\le k \le K \le 2^{n_b - 1}$, the most significant qubit of $|k - 1\rangle_b$ is $0$ and the output is 
    \[|k-1\rangle_b|0\rangle_a \HH_p|\psi\rangle_s + |g\rangle.\]
    In conclusion, the circuit shown in \Cref{fig:Tpp} correctly realizes $\T_{p,p}$.
    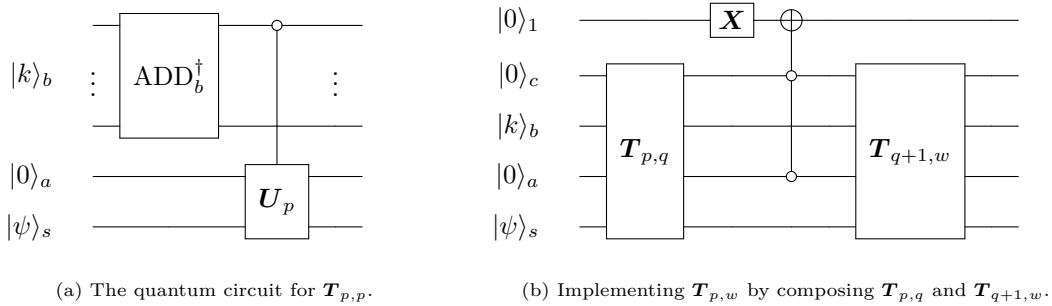
\begin{figure}
        \centering
        \begin{subfigure}{0.45\textwidth}
            \centering
            \[\Qcircuit @C=1em @R=1em {
                \lstick{}& & \multigate{2}{\text{ADD}^{\dagger}_b} & \ctrlo{3} & \qw & \qw \\
                \lstick{|k\rangle_b}& \vdots & \nghost{\text{ADD}^{\dagger}_b} &  & \vdots  \\
                \lstick{}& & \ghost{\text{ADD}^{\dagger}_b} & \qw & \qw & \qw\\
                \lstick{|0\rangle_a}& & \qw & \multigate{1}{\U_p} & \qw & \qw\\
                \lstick{|\psi \rangle_s}& & \qw & \ghost{\U_p} & \qw & \qw\\
            }\]
            \caption{The quantum circuit for $\T_{p, p}$.}
            \label{fig:Tpp}
        \end{subfigure}
        \begin{subfigure}{0.45\textwidth}
            \centering
            \[\Qcircuit @C=1em @R=1em {
                \lstick{|0\rangle_1}& & \qw & \gate{\X} & \targ & \qw & \qw & \qw  & \qw\\
                \lstick{|0\rangle_c}& & \multigate{3}{\T_{p, q}} & \qw & \ctrlo{-1} & \qw & \multigate{3}{\T_{q+1, w}} & \qw  & \qw\\
                \lstick{|k\rangle_b}& & \ghost{\T_{p, q}} & \qw & \qw & \qw & \ghost{\T_{q+1, w}} & \qw  & \qw\\
                \lstick{|0\rangle_a}& & \ghost{\T_{p, q}} & \qw & \ctrlo{-2} & \qw & \ghost{\T_{q+1, w}} & \qw  & \qw\\
                \lstick{|\psi \rangle_s}& & \ghost{\T_{p, q}} & \qw & \qw & \qw & \ghost{\T_{q+1, w}} & \qw  & \qw\\
            }\]
            \caption{Implementing $\T_{p, w}$ by composing $\T_{p, q}$ and $\T_{q+1, w}$.}
            \label{fig:TTtT}
        \end{subfigure}
        \caption{Illustration of the two main steps for constructing $\T_{1, K}$. (a) base case and (b) recursive composition.}
        \label{fig:TTT}
    \end{figure}
    \item \textbf{Recursive composition:} Given circuits for $\T_{p, q}$ and $\T_{q+1, w}$, we can construct $\T_{p, w}$ as shown in \Cref{fig:TTtT}.
    For $1 - p\le k \le K$, applying this composition to $|0\rangle_c |k\rangle_b |0\rangle_a |\psi\rangle_s$ gives
    \[\begin{aligned}
        |0\rangle_c |k\rangle_b |0\rangle_a |\psi\rangle_s & \longrightarrow |0\rangle_1 \left(|0\rangle_c |k - q + p - 1\rangle_b |0\rangle_a \left(\prod_{j = p}^{\min\{q, p + k - 1\}}\HH_j\right) |\psi\rangle_s + | g \rangle\right)\\
        & \longrightarrow |0\rangle_1 |0\rangle_c |k - q + p - 1\rangle_b |0\rangle_a \left(\prod_{j = p}^{\min\{q, p + k - 1\}}\HH_j\right) |\psi\rangle_s + |1\rangle_1 |g \rangle\\
        & \longrightarrow |0\rangle_1 \left(|0\rangle_c |k - w + p - 1\rangle_b |0\rangle_a \left(\prod_{i = q+1}^{\min\{w, p + k - 1\}}\HH_i \prod_{j = p}^{\min\{q, p + k - 1\}}\HH_j\right) |\psi\rangle_s + |g'\rangle \right) + |1\rangle_1 |g''\rangle \\
        & = |0\rangle_1|0\rangle_c |k - w + p - 1\rangle_b |0\rangle_a \left(\prod_{j = p}^{\min\{w, p + k - 1\}}\HH_j\right) |\psi\rangle_s + \left(|0\rangle_1|g'\rangle + |1\rangle_1 |g''\rangle \right),
    \end{aligned}\]
    where $|0\rangle_1|g'\rangle + |1\rangle_1|g''\rangle$ is orthogonal to the subspace $|0\rangle_1 |0\rangle_c \otimes \I_b \otimes |0\rangle_a\otimes \I_s$, verifying that the circuit implements $\T_{p, w}$. This recursive composition requires one additional ancilla qubit and one multi-controlled NOT gate, which can be implemented using $O(n_a + n_c)$ primitive gates with a few ancillas.
    \item \textbf{Full construction:} Let $n_b = \left\lceil \log_2 K \right\rceil + 1$. Starting from $\T_{p,p}$ circuits, we can construct $\T_{1, K}$ using $\left\lceil \log_2 K \right\rceil$ ancilla qubits. Thus, $n_c$ can be chosen as $\left\lceil \log_2 K \right\rceil$. The overall cost is one query to each controlled-$\U_k$ and $O(K(\log K + n_a))$ additional primitive quantum gates. 
\end{enumerate}
\subsection{Block-encoding matrices with Kronecker-sum structure}
It is often the case that matrices of practical interest exhibit a Kronecker-sum structure, meaning they can be expressed as a sum of Kronecker products. By identifying and exploiting this structure, together with our method for implementing Kronecker products, the task of block-encoding a complicated matrix can be reduced to block-encoding several simpler matrices. We demonstrate this approach through several representative examples.
\subsubsection{Generalized Sylvester equation} 
The generalized Sylvester equation, typically expressed as
\[\sum_{k=1}^m \A_k \X \B_k = \C,\]
where $\A_k, \B_k$ and $\C$ are known matrices and $\X$ is unknown, arises naturally in a wide range of scientific and engineering applications \cite{doi:10.1137/130912839}. Exploiting the vectorization identity
\[\texttt{vec}(\A\X\B) = (\B^T \otimes \A) \texttt{vec}(\X), \]
the equation can be reformulated as the linear system
\[\left(\sum_{k = 1}^m \B_k^T \otimes \A_k\right) \texttt{vec}(\X) = \texttt{vec}(\C).\]
Given block-encodings of $\A_k$, $\B_k$, and $\C$, one can first construct block-encodings of $\B_k^T\otimes \A_k$'s, and then apply the LCU method \cite{10.1145/3313276.3316366} to obtain the block-encoding of 
\[\sum_{k=0}^m \B_k^T \otimes \A_k.\]
Further, the block-encoding of $\texttt{vec}(\C)$ can be constructed directly from the block-encoding of $\C$, as described in \Cref{subsubsec:vec}. Equipped with these block-encodings, the generalized Sylvester equation can then be solved on a quantum computer using QLSAs.
\subsubsection{Discretised differential operators} 
Differential equations are widely used to model physical processes. As the dimension increases, the size of the discretised matrix grows exponentially. However, many such matrices can be expressed as a sum of Kronecker products, which makes them well-suited for efficient block-encoding. 

As a concrete example, consider the problem of pricing multi-asset derivatives by solving the Black-Scholes partial differential equation (PDE) using the finite-difference method. This task suffers from the curse of dimensionality, namely the exponential growth of computational complexity with the number of underlying assets. In \cite{9618807}, the authors present a quantum algorithm for this problem. By discretising the spatial variables in the Black-Scholes PDE, the problem is reformulated as an ODE system:
\[\frac{\mathrm{d} }{\mathrm{d} \tau} \y(\tau) = \F \y(\tau) + \cc(\tau),\]
where $\cc(\tau)$ encodes the inhomogeneous contributions from boundary conditions, and the coefficient matrix $\F$ admits a Kronecker-sum structure:
\[\begin{aligned}
    \F & := \F^{2 \mathrm{nd}} + \F^{1 \mathrm{st}}, \\
    \F^{2 \mathrm{nd}} & :=\sum_{i=1}^d \frac{\sigma_i^2}{2 h_i^2} \I^{\otimes i-1} \otimes \D^{2 \mathrm{nd}} \otimes \I^{\otimes d-i} + \sum_{i=1}^{d-1} \sum_{j=i+1}^d \frac{\sigma_i \sigma_j \rho_{i j}}{4 h_i h_j} \I^{\otimes i-1} \otimes \D^{1 \mathrm{st}} \otimes \I^{\otimes j-i-1} \otimes \D^{1 \mathrm{st}} \otimes \I^{\otimes d-j} ,\\
    \F^{1 \mathrm{st}} & :=\sum_{i=1}^d \frac{1}{2 h_i}\left(r-\frac{1}{2} \sigma_i^2\right) \I^{\otimes i-1} \otimes \D^{1 \mathrm{st}} \otimes \I^{\otimes d-i},
\end{aligned}\]
where $d$ is the number of assets, $\sigma_i, h_i, \rho_{ij}, r$ are given real parameters, $\I$ denotes the identity matrix. The one-dimensional finite-difference matrices for first- and second-order derivatives are
\[\D^{1 \mathrm{st}}:=\left(\begin{array}{cccccc}
    0 & 1 & & & & \\
    -1 & 0 & 1 & & & \\
    & -1 & 0 & 1 & & \\
    & & \ddots & \ddots & \ddots & \\
    & & & -1 & 0 & 1 \\
    & & & & -1 & 0
\end{array}\right),\quad  \D^{2 \mathrm{nd}}:=\left(\begin{array}{cccccc}
    -2 & 1 & & & & \\
    1 & -2 & 1 & & & \\
    & 1 & -2 & 1 & & \\
    & & \ddots & \ddots & \ddots & \\
    & & & 1 & -2 & 1 \\
    & & & & 1 & -2
\end{array}\right).\]

To solve such ODEs on a quantum computer, the state-of-the-art approaches rely on efficient block-encodings of the coefficient matrix \cite{berry2017quantum,an2023quantumalgorithmlinearnonunitary,jin2024schrodingerizationbasedquantumalgorithms}. The block-encodings of $\D^{1 \mathrm{st}}$ and $\D^{2 \mathrm{nd}}$ can be efficiently implemented by the method of \cite{camps2024explicit}. Once the two block-encodings are available, the block-encoding of $\F$ follows naturally from its Kronecker-sum structure, combined with the LCU method. 
\subsubsection{Adjacent matrix of extended binary tree.}
\begin{figure}
    \centering
    \begin{subfigure}{0.45\textwidth}
        \centering
        \[\Qcircuit @C=1em @R=1em {
            \lstick{|0\rangle}& \multigate{1}{\PP^\dagger}  & \qw & \qw & \qw & \qw &\\
            \lstick{|j_2\rangle}& \ghost{\PP^{\dagger}}  & \multigate{1}{\PP^\dagger}& \qw & \qw & \qw &\\
            \lstick{|j_1\rangle}& \qw  & \ghost{\PP^{\dagger}}& \multigate{1}{\PP^\dagger} & \qw & \qw &\\
            \lstick{|j_0\rangle}&   \qw & \qw & \ghost{\PP^{\dagger}} & \gate{\HH} & \qw & 
        }\]
        \caption{Quantum circuit of $\U_1$.}
        \label{fig:b_kron_c}
    \end{subfigure}
    \begin{subfigure}{0.45\textwidth}
        \centering
        \[\Qcircuit @C=1em @R=1.3em {
            \lstick{|0\rangle}& \gate{R_y(\theta_0)} & \gate{R_y(\theta_1)} & \gate{R_y(\theta_2)} & \qw\\
            \lstick{|j_2\rangle}& \ctrlo{-1}& \ctrl{-1} & \ctrlo{-1} & \qw\\
            \lstick{|j_1\rangle}& \qw & \qw & \ctrlo{-1} & \qw\\
            \lstick{|j_0\rangle}& \qw & \qw & \ctrlo{-1} & \qw
        }\]
        \caption{Quantum circuit of $\U_2$.}
        \label{fig:bc_diag}
    \end{subfigure}
    \caption{Block-encodings of the three matrices in the decomposition \Cref{eq:deco_bin}. Subfigure (a) shows the block-encoding of the first matrix, whose inverse gives that of the second matrix, while subfigure (b) shows the block-encoding of the third matrix.}
    \label{fig:placeholder}
\end{figure}
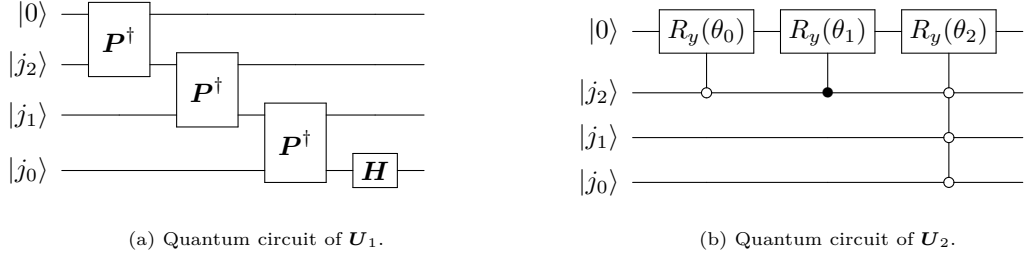

Consider the adjacent matrix of an extended binary tree, given by  
\begin{equation}\label{eq:bin_tree}
    \A = \left[\begin{array}{llllllll}
        \gamma & \beta & & & & & & \\
        \beta & \alpha & \beta & \beta & & & & \\
        & \beta & \alpha & & \beta & \beta & & \\
        & \beta & & \alpha & & & \beta & \beta \\
        & & \beta & & \gamma & & & \\
        & & \beta & & & \gamma & & \\
        & & & \beta & & & \gamma & \\
        & & & \beta & & & & \gamma
    \end{array}\right],
\end{equation}
where $\alpha, \beta, \gamma \in(0,1)$. Block-encodings of such matrices were analyzed in \cite{camps2024explicit}. Here we show that a more efficient block-encoding can be obtained by expressing $\A$ as a sum of Kronecker products. Specifically, we decompose $\A$ as 
\begin{equation}\label{eq:deco_bin}
    \begin{aligned}
        \A &= \left[\begin{array}{llllllll}
            \beta & & & & & & & \\
            \beta & & & & & & & \\
            & \beta & & & & & & \\
            & \beta & & & & & & \\
            & & \beta & & & & & \\
            & & \beta & & & & & \\
            & & & \beta & 0& 0& 0& 0\\
            & & & \beta & 0& 0& 0& 0
        \end{array}\right] + \left[\begin{array}{llllllll}
            \beta & \beta & & & & & & \\
            & & \beta & \beta & & & & \\
            & & & & \beta & \beta & & \\
            & & & & & & \beta & \beta \\
            & & & & & &0 &0 \\
            & & & & & &0 &0 \\
            & & & & & &0 &0 \\
            & & & & & &0 &0
        \end{array}\right] + \left[\begin{array}{llllllll}
            \gamma - 2\beta&&&&&&& \\ & \alpha &&&&&&\\ && \alpha&&&&& \\ &&& \alpha &&&&\\ &&&&\gamma &&&\\ &&&&&\gamma&&\\ &&&&&&\gamma&\\ &&&&&&&\gamma
        \end{array}\right].
    \end{aligned}
\end{equation}
The first matrix in \Cref{eq:deco_bin} can be written as a Kronecker product:
\[\sqrt{2}\beta \begin{bmatrix}1&0&0&0&0&0&0&0\\0&1&0&0&0&0&0&0\\0&0&1&0&0&0&0&0\\0&0&0&1&0&0&0&0\end{bmatrix}\otimes \begin{bmatrix}
    \frac{1}{\sqrt{2}}\\ \frac{1}{\sqrt{2}}\end{bmatrix} =:\sqrt 2\beta  \B\otimes \C.
\]
By \Cref{def:bc}, the block-encodings of $\B$ and $\C$ can be choosen as
\[\I_8 = \begin{bmatrix}
1&0&0&0&0&0&0&0\\0&1&0&0&0&0&0&0\\0&0&1&0&0&0&0&0\\0&0&0&1&0&0&0&0\\0&0&0&0&1&0&0&0\\0&0&0&0&0&1&0&0\\0&0&0&0&0&0&1&0\\0&0&0&0&0&0&0&1
\end{bmatrix}, \quad \text{and}\quad \HH = \frac1{\sqrt2} \begin{bmatrix}
    1 & 1\\1 & -1
\end{bmatrix}.\]
To convert $\I_8\otimes \HH$ into a block-encoding of $\B\otimes \C$, we apply \Cref{lem:move} and obtain the permutation matrix 
\[\PP = \begin{bmatrix}1&0&0&0\\0&0&1&0\\0&0&0&1\\0&1&0&0\end{bmatrix},\]
which can be implemented with two CNOT gates. Consequently, the unitary
\[ \U_1 := (\I_8\otimes \HH) \cdot \left(\I_4\otimes \PP^\dagger\right) \cdot \left(\I_2\otimes \PP^\dagger \otimes \I_2\right)\cdot \left(\PP^\dagger \otimes \I_4\right)\]
serves as a block-encoding of $\B\otimes \C$, and the corresponding quantum circuit is shown in \Cref{fig:b_kron_c}.
To verify, note that $\PP^\dagger |0 b\rangle = |b0\rangle$, for all $b\in \{0, 1\}$. Applying the three $\PP^\dagger$ layers followed by $\HH$ gives
\[\begin{aligned}
    |0j_2j_1j_0\rangle & \longrightarrow |j_20j_1j_0\rangle \longrightarrow |j_2j_10j_0\rangle \longrightarrow |j_2j_1j_00\rangle\\
    & \longrightarrow \frac{1}{\sqrt{2}}|j_2j_1j_0\rangle \left(|0\rangle + |1\rangle \right).
\end{aligned}\]
Taking the overlap with $|0i_2i_1i_0\rangle $ yields
\[ \langle 0i_2i_1i_0|\U_1|0 j_2j_1j_0 \rangle = \begin{cases}
    0, & j_2 = 1,\\
    \frac1{\sqrt2}, & j_2 = 0 \text{ and } (i_2, i_1) = (j_1, j_0),\\
    0, & \text{otherwise.}
\end{cases} \]
Equivalently, with $j:= 4j_2 + 2j_1 + j_0$ and $i := 4i_2 + 2i_1 + i_0$, the nonzero entries appear precisely when $j_2 = 0$ and $i\in \{2j, 2j + 1\}$, each with value $1/\sqrt{2}$. This exactly reproduces the structure of $\B\otimes \C$. 

In the same way, a block-encoding of the second matrix in \Cref{eq:deco_bin} can be obtained either by direct construction or by taking the inverse of $\U_1$. Hence, block-encodings of the first two terms in the decomposition are established. 

\begin{figure}
    \centering
    \[\Qcircuit @C=1em @R=1em {
        \lstick{|0\rangle_1}& \gate{Z} & \gate{R_y(\zeta)} & \ctrlo{1} & \ctrlo{1} & \ctrl{2} & \gate{Z} & \gate{R_y(\zeta)}& \qw\\
        \lstick{|0\rangle_1}& \qw & \gate{H} & \ctrlo{1} & \ctrl{1} & \qw & \gate{H} & \qw& \qw\\
        \lstick{|0\rangle_1}& \qw & \qw & \multigate{1}{U_1}& \multigate{1}{U_1^\dagger} & \multigate{1}{U_2} & \qw & \qw & \qw \\
        \lstick{|j\rangle_3}& \qw & \qw & \ghost{U_1} & \ghost{U_1^\dagger} & \ghost{U_2} & \qw & \qw & \qw
    }\]
    \caption{Block-encoding of the adjacent matrix of an extended binary tree.}
    \label{fig:bin_bc}
\end{figure}
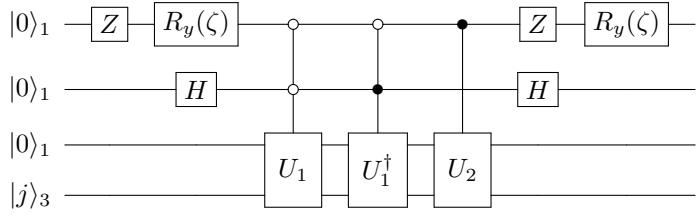

For the third term in \Cref{eq:deco_bin}, which is diagonal, we employ the block-encoding circuit $\U_2$ shown in \Cref{fig:bc_diag}. Define 
\[\widetilde \alpha := \cos \frac{\theta_0}{2}, \quad \widetilde \gamma := \cos \frac{\theta_1}{2},\quad  \widehat \gamma := \cos \frac{\theta_0 + \theta_2}{2}, \]
then the top-left $8\times 8$ block of $\U_2$ is 
\[\text{diag}\left(\widehat \gamma, \widetilde \alpha, \widetilde \alpha, \widetilde \alpha, \widetilde \gamma, \widetilde \gamma, \widetilde \gamma, \widetilde \gamma\right).\]
The parameters $\widetilde \alpha$, $\widetilde \gamma$, $\widehat \gamma$  will be specified later. Next, we perform LCU on the block-encodings of the three matrices, obtaining 
\[  x \B\otimes \C +  x \B^T \otimes \C^T + \left(1 - 2x\right) \text{diag}\left(\widehat \gamma, \widetilde \alpha, \widetilde \alpha, \widetilde \alpha, \widetilde \gamma, \widetilde \gamma, \widetilde \gamma, \widetilde \gamma\right) = \frac1c \A. \]
Here, the parameters $x, c, \widehat \gamma, \widetilde \alpha, \widetilde \gamma$ are chosen such that 
\begin{equation}\label{eq:assign}
    \begin{cases}
    \frac x{\sqrt2} = \frac{\beta}c,\\
    (1-2x)\widehat \gamma + \sqrt 2 x = \frac{\gamma}{c},\\
    (1-2x)\widetilde \gamma = \frac{\gamma}{c},\\
    (1-2x)\widetilde \alpha = \frac{\alpha}{c}.
    \end{cases} \Longrightarrow \begin{cases}x = \frac{\sqrt 2 \beta}{c},\\ \widehat \gamma = \frac{\gamma - 2\beta}{c - 2\sqrt{2}\beta},\\ \widetilde \gamma = \frac{\gamma}{c - 2\sqrt2 \beta},\\ \widetilde\alpha = \frac{\alpha}{c - 2\sqrt2 \beta}.\end{cases}
\end{equation}
To implement this procedure on a quantum computer, the parameters must further satisfy
\[0\le x \le \frac12, \quad |\widehat \gamma|, |\widetilde \alpha|, |\widetilde \gamma| \le 1.\]
By choosing $c = 2\sqrt 2 + 2$, these conditions can be fulfilled since $\alpha, \beta, \gamma \in (0, 1)$. For explicitly given values of $\alpha, \beta, \gamma$, one may select a smaller $c$. Setting $\zeta = 2\arccos\sqrt{2x}$, we obtain 
\[\begin{aligned}
    \left[R_y(\zeta)\cdot Z\right]\otimes H &= \begin{bmatrix}
        \sqrt{2x}& \sqrt{1-2x}\\\sqrt{1-2x}& -\sqrt{2x}
    \end{bmatrix}\otimes \begin{bmatrix}\frac{\sqrt 2}{2}& \frac{\sqrt2}{2}\\\frac{\sqrt 2}{2}& -\frac{\sqrt2}{2}\end{bmatrix} = \begin{bmatrix}
        \sqrt{x}&\sqrt{x}&\sqrt{\frac12-x}&\sqrt{\frac12 - x}\\
        \sqrt{x}&*&*&*\\
        \sqrt{\frac12-x}&*&*&*\\
        \sqrt{\frac12-x}&*&*&*
    \end{bmatrix}.
\end{aligned} \]
In summary, the quantum circuit for the block-encoding of $\frac1c A$ is given in \Cref{fig:bin_bc}. 
This construction naturally generalizes to extended binary trees of greater depth. Suppose the tree has depth $n$. Then $A$ can be block-encoded using $O(n^2)$ quantum gates and $3$ ancillary qubits. The $O(n^2)$ complexity arises from the multi-controlled rotation $R_y(\theta_2)$ required in block-encoding the diagonal part, while the remaining components of the block-encoding can be implemented with only $O(n)$ gates. 

In the special case $\beta\in (0, 1/3)$ with $\alpha = 1- 3\beta$ and $\gamma = 1- \beta$, i.e., when $\A$ is a symmetric stochastic matrix, setting $c = 1$ suffices, and the block-encoding of $\A$ becomes exact. Moreover, in this regime we have $\widehat \gamma = \widetilde \alpha$, which allows us to eliminate the final multi-controlled rotation $R_y(\theta_2)$ in \Cref{fig:bc_diag}. The overall gate complexity of the block-encoding then reduces to $O(n)$.

In contrast, the construction of \cite{camps2024explicit} requires $5$ ancilla qubits and a more involved circuit to obtain a block-encoding with normalization $c = 8$. Our construction uses only $3$ ancillas and achieves a strictly better constant: in general we can choose $c \le 2\sqrt 2 + 2 < 5$, and in favorable parameter regimes even $c = 1$. This demonstrates the benefit of identifying and exploiting the Kronecker-sum structure. 

\section{Conclusion}\label{sec:4}
In this work we developed new techniques for implementing matrix products within the block-encoding framework, extending the scope and efficiency of existing constructions. First, we addressed the general case where the matrices are not necessarily square and their dimensions are not restricted to powers of two, thereby broadening the applicability of block-encoded operations. Second, we introduced a qubit-efficient method for realizing matrix-matrix multiplications that significantly reduces ancilla requirements at the cost of a modest increase in gate complexity. This improvement is particularly pronounced in the sequential product setting, where the number of required ancilla qubits decreases exponentially. In addition, we refined the implementation of the Kronecker product compared to prior work \cite{PhysRevA.102.052411}, replacing SWAP operations with pairs of CNOT gates and thus lowering gate cost. By applying the same qubit-efficient principle from our matrix-matrix multiplication construction, we also developed a qubit-efficient Kronecker product implementation. Moreover, we rederived the Hadamard product between block-encodings and extend its construction to the convolution operation and vectorization operator. Finally, we demonstrated the utility of these basic product operations in applications such as time-dependent Hamiltonian simulation and the block-encoding of the adjacent matrix of an extended binary tree, where our constructions yield concrete resource reductions relative to existing approaches.

Taken together, our results enrich the toolbox for manipulating block-encodings, offering both qubit efficiency and structural flexibility. Looking ahead, it will be of interest to integrate these techniques into practical quantum algorithms---such as Hamiltonian simulation and quantum optimization---and to further explore linear operations and their applications in emerging areas of quantum computational science. 


\appendix
\section{Qubit-efficient implementation of Kronecker product}\label{app:A}
Similarly to the matrix-matrix product, the Kronecker product between block-encoded matrices admits a qubit-efficient implementation.

\begin{figure}
    \centering
    \[\Qcircuit @C=1em @R=1em{
        \lstick{|0\rangle_1}& \qw & \qw & \qw & \qw & \qw \barrier[-2em]{7} & \gate{\X} & \targ \barrier[1em]{7}& \qw & \qw & \qw & \qw & \qw & \qw\\
        \lstick{|0\rangle_{c-t-b + s}}& \qw & \qw & \qw & \qw & \qw & \qw & \qw & \qw & \qw& \qw& \multigate{1}{\U_{\C}} & \qw& \qw\\
        \lstick{|0\rangle_{b-s}}& \qw & \multigate{1}{\U_{\B}}& \qw & \qw & \qw & \qw & \ctrlo{-2} & \qw & \qw& \qw& \ghost{\U_{\C}} & \qw& \qw\\
        \lstick{|x\rangle_s}& \qw & \ghost{\U_{\B}} & \qw & \qw & \qw & \qw & \qw & \qw & \qw& \qw& \qw & \qw& \qw\\
        \lstick{|y\rangle_t}& \qw & \qw & \qw & \qw & \qw  & \qw & \qw & \qw & \qw& \qw& \sgate{\U_{\C}}{-2} & \qw& \qw\\
        &&\dstick{\parbox{3cm}{\centering Block-encoding\\ of $\B\otimes \I_{2^c}$}}&&&& \dstick{\parbox{3cm}{~ ~ ~ ~ ~ ~ Permutation}} &&&&&\dstick{\parbox{3cm}{\centering ~ ~ Block-encoding\\ ~ ~ of $\I_{2^b}\otimes \C$}}&&\\
        &&&&&&&&\\
        &&&&&&&&\\
        } 
    \]
    \caption{Quantum circuit for qubit-efficient implementation of the Kronecker product. Two barriers divide the circuit into three phases. The left and right phases implement the block-encodings of $\B\otimes \I_{2^c}$ and $\I_{2^b}\otimes \C$, respectively. The middle phase applies the permutation used in Example \ref{ex:1}.}
    \label{fig:qubit_saving_kron}
\end{figure}
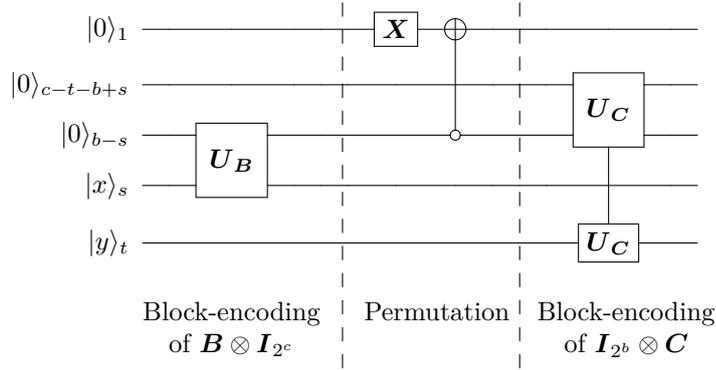

For simplicity, consider first the case where $\B \in \mathbb C^{2^{\scriptstyle{s}}\times 2^{\scriptstyle{s}}}$ and $\C \in \mathbb C^{2^{\scriptstyle{t}}\times 2^{\scriptstyle{t}}}$ are square matrices whose dimensions are powers of two, and let their block-encodings be $\U_{\B}\in \mathbb C^{2^{\scriptstyle{b}}\times 2^{\scriptstyle{b}}}$ and $\U_{\C}\in \mathbb C^{2^{\scriptstyle{c}}\times 2^{\scriptstyle{c}}}$. Without loss of generality, assume that $\U_{\C}$ uses more ancilla qubits, i.e., $c-t > b-s$. Noting that 
\[\B\otimes \C = (\I_{2^b} \otimes \C) \left(\B \otimes \I_{2^c}\right),\]
and the block-encodings of $\I_{2^b}\otimes \C$ and $\B\otimes \I_{2^c}$ are obtained directly from $\U_{\C}$ and $\U_{\B}$, respectively, we may apply the qubit-efficient matrix-matrix product construction to realize the block-encoding of $\B\otimes \C$. The resulting quantum circuit is presented in \Cref{fig:qubit_saving_kron}; it implements a block-encoding of $\B\otimes \C$ while requiring only one additional ancilla qubit beyond those used by $\U_{\B}$ and $\U_{\C}$.
The effect of this circuit on the input state $|0\rangle_1 |0\rangle_{c-t} |x\rangle_s|y\rangle_t$ is as follows:
\[\begin{aligned}
    |0\rangle_1 |0\rangle_{c-t} |x\rangle_s|y\rangle_t & \longrightarrow |1\rangle_1 \left(|0\rangle_{c-t} \B|x\rangle_b |y\rangle_c + |g\rangle\right)\\
    &\longrightarrow |0\rangle_1 |0\rangle_{c-t} \B|x\rangle_b |y\rangle_c + |1\rangle_1 |g\rangle\\
    &\longrightarrow |0\rangle_1 |0\rangle_{c-t} \B|x\rangle_b \C|y\rangle_c + |0\rangle_1 |g'\rangle + |1\rangle_1 |g''\rangle,
\end{aligned}\]
where both of the last two terms are orthogonal to the subspace $|0\rangle_{c-t+1}\otimes \I_{2^{s+t}}$. Consequently, the circuit implements a block-encoding of $\B\otimes \C$ using only one additional ancilla qubit. 

For the general case where $\B$ and $\C$ have arbitrary sizes, the same construction applies after embedding $\B$ and $\C$ into larger $2^s\times 2^s$ and $2^t\times 2^t$ blocks. The output registers $|\cdot\rangle_s$ and $|\cdot\rangle_t$ can then be rearranged as in \Cref{subsec:kron_prod} so that $\B\otimes \C$ occupies the top-left corner of the overall unitary. 

\bibliographystyle{elsarticle-num} 
\bibliography{mybib}






\end{document}